\documentclass[10pt]{iopart}
\usepackage{cite}
\usepackage{graphicx}

\begin{document}

\title[Strontium optical lattice clocks]{Strontium optical lattice clocks for practical realization of the metre and secondary representation of the second}

\author{Marcin Bober, Piotr Morzy\'nski, Agata Cygan, Daniel Lisak, Piotr Mas\l{}owski, Mateusz Prymaczek, Piotr Wcis\l{}o, Piotr Ablewski, Mariusz Piwi\'nski, Szymon W\'ojtewicz, Katarzyna Bielska, Dobros\l{}awa Bartoszek-Bober,  { Ryszard S. Trawi\'nski}, Micha\l{} Zawada, Roman Ciury\l{}o}

\address{Institute of Physics, Faculty of Physics, Astronomy and Informatics, Nicolaus Copernicus University, Grudzi\c{a}dzka 5, PL-87-100 Toru\'n, Poland}

\author{Jerzy Zachorowski, Marcin Piotrowski, Wojciech Gawlik}
\address{M. Smoluchowski Institute of Physics, Faculty of Physics, Astronomy and Informatics, Jagiellonian University, St. \L{}ojasiewicza 11, PL-30-348 Krak\'ow, Poland}

\author{Filip Ozimek, Czes\l{}aw Radzewicz}
\address{Institute of Experimental Physics, Faculty of Physics, University of Warsaw, Pasteura 5, PL-02-093 Warsaw, Poland}

\ead{zawada@fizyka.umk.pl}

\begin{abstract}
We present a system of two independent strontium optical lattice standards probed with a single shared ultra-narrow laser. The absolute frequency of the clocks can be verified by the use of Er:fiber  optical frequency comb  with the GPS-disciplined Rb frequency standard. We report hertz-level spectroscopy of the clock line and measurements of frequency stability of the two  strontium optical lattice clocks.

\end{abstract}

\vspace{2pc}

\submitto{\MST}  doi:10.1088/0957-0233/26/7/075201
\maketitle

\ioptwocol

\section{Introduction}
Ultracold neutral atoms in an optical lattice \cite{Ido03} and trapped single-ions \cite{Rosenband08}  are two well-known approaches for development of optical frequency standards.

The ${}^{1}S_{0}$ -- ${}^{3}P_{0}$ transition in neutral strontium was recommended by the International Committee for Weights and Measures for practical realization of the metre and secondary representation of the second.  Although best realisations of the strontium atomic clocks reached accuracy and stability at the $10^{-17}$ level or better \cite{Ye14, LeTargat13, Hinkley13, Falke14, Ushijima14}, the International Bureau of Weights and Measures (BIPM) set practical relative uncertainties above $1\times 10^{-15}$ in case of fermionic isotope ${}^{87}$Sr \cite{BIPM87} and $1\times 10^{-14}$ in case of bosonic isotope ${}^{88}$Sr \cite{BIPM88}. 

This conservatism stems from the fact that to calculate the recommended frequency values the BIPM uses the weighted average of independently obtained frequencies and the pool of available strontium optical clocks worldwide is still small.
Enlarging this pool is an essential prerequisite for a possible redefinition of the second.

The optical clocks are not only promising candidates for the future primary standards, but serve as  sensitive probes in such areas as searches for variations
of fundamental constants \cite{Rosenband08,Peik04,Fortier07,Blatt08,Peik10}, relativistic
geodesy  \cite{Bjerhammar85,Delva13}, tests of relativity \cite{Schiller09,Chou10} and searches for dark matter \cite{Derevianko14}.

We developed a system of two independent strontium optical lattice clocks. The system consists of two atomic standards interrogated by a shared ultra-narrow laser, pre-stabilised  to  a high-Q optical cavity and an optical frequency comb. 
We demonstrate hertz-level spectroscopy of the clock line and measurements of frequency stability of the standards.

\section{Optical lattice standards}

\begin{figure}[ht]
\centering
\includegraphics[width=1\columnwidth]{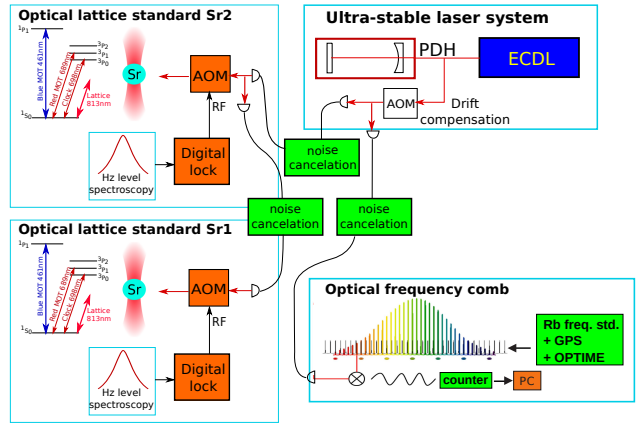}
\caption{(Color online) A simplified scheme of the system of two optical lattice clocks Sr1 and Sr2.  The clouds of atoms in Sr1 and Sr2 are independently probed by an ultrastable laser.  The frequencies of both beams are locked to the narrow resonances in each standard by a digital lock and acousto-optic frequency shifters. The frequency of the clock transition can be compared with a GPS-disciplined Rb frequency standard by the use of an optical frequency comb. 
 \label{fig:general}}
\end{figure}

A simplified scheme of the system of two optical lattice clocks is depicted in Fig. \ref{fig:general}.
The two optical frequency standards (Sr1 and Sr2) are based on the  ${}^{1}S_{0}$ -- ${}^{3}P_{0}$ transition in neutral strontium atoms. The Sr1 can operate with bosonic  isotope ${}^{88}$Sr while Sr2 can operate with either bosonic  ${}^{88}$Sr or fermionic  ${}^{87}$Sr isotope. 
Two clouds of atoms in Sr1 and Sr2 are independently probed by an ultrastable laser with spectral width below 1~Hz. 
The laser beam is split into two optical paths.
The frequencies of both beams are independently locked to the narrow atomic resonances in each standard.
The digital lock gives the feedback to the  acousto-optic frequency shifters.  
The frequency of the clock transition can be compared with a GPS-disciplined Rb frequency standard by the use of an optical frequency comb.

In both standards  atoms are tightly confined in a Lamb-Dicke regime in an optical lattice formed by a standing wave inside an optical cavity. The Lamb-Dicke regime \cite{Dicke53} effectively suppress all motional effects preventing any Doppler shifts of the measured transition. The lattice is operated at the so-called magic wavelength \cite{Takamoto05,Ye08} when the light shifts of both ${}^{1}S_{0}$ and ${}^{3}P_{0}$ energy levels compensates each others with high accuracy. 
Moreover, the 1D lattice is oriented vertically.  In such orientation tunnelling of atoms between lattice sites is prevented because of small energy differences produced by the Earth's gravity field between levels in adjacent lattice sites.
Since the depth of the optical trap is relatively low, the trap can capture atoms with temperatures significantly below  1~mK. Strontium atoms on the other hand have low vapour pressure and need to be heated in an atomic oven to $500^{\circ}$C to form an atomic beam. Therefore, a system of a Zeeman slower and two stages magneto-optical trap is needed  to load the atoms into the  optical lattices.
Details of the system are described in the following sections.

\subsection{Vacuum chambers}

\begin{figure}[hb]
\centering
\includegraphics[width=0.9\columnwidth]{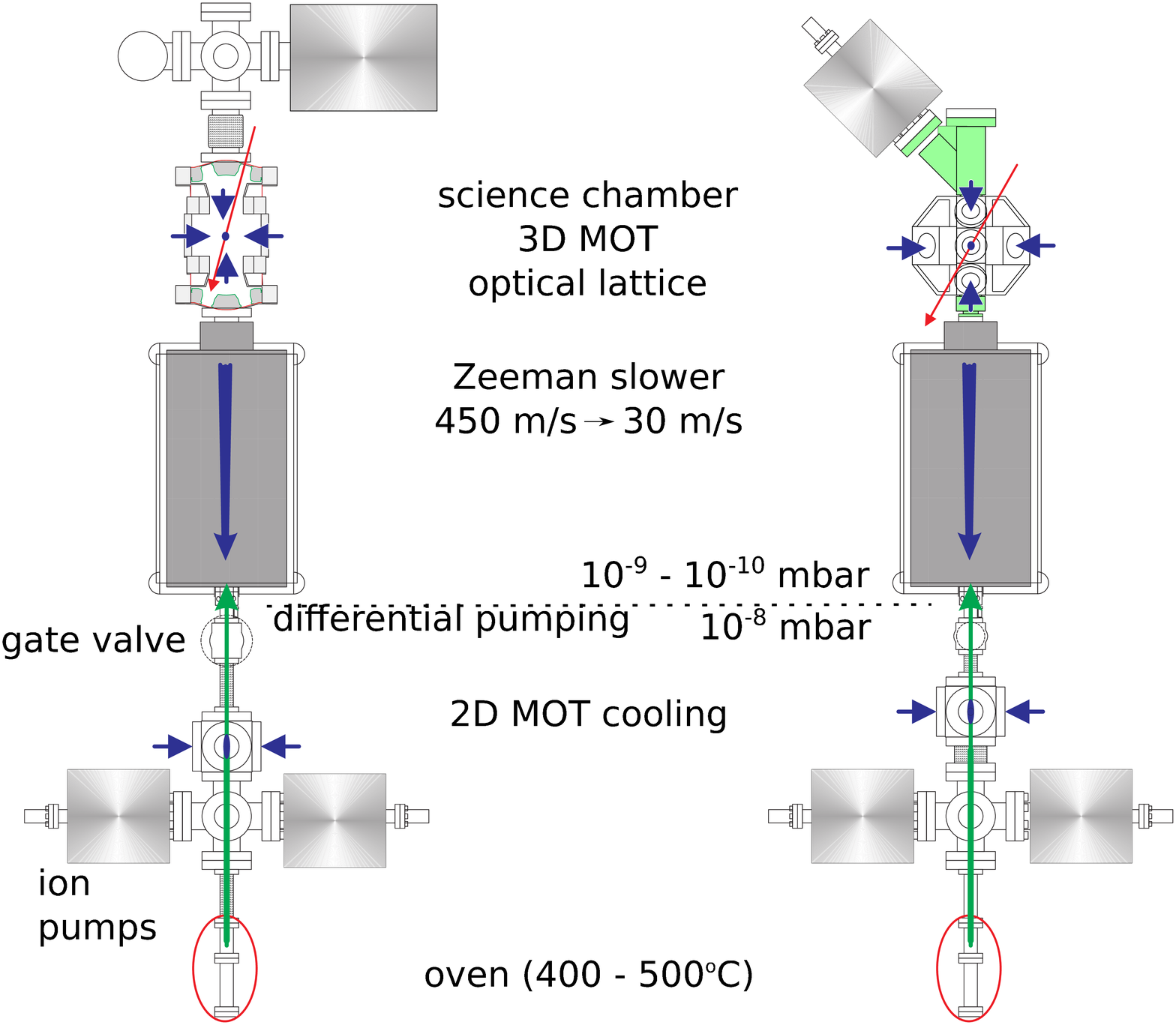}
\caption{(Color online) The vacuum set-ups of Sr1 and Sr2. Blue and red arrows indicate laser beams, green arrows are the atomic beams. \label{fig:vac}}
\end{figure}

The vacuum set-ups of Sr1 and Sr2 are presented in Fig. \ref{fig:vac}. The long trapping lifetime in the optical lattice require ultra-high vacuum condition. Therefore, the vacuum system is split into two parts
connected by a long, thin tube providing efficient differential pumping.  The first one contains the atomic oven producing collimated strontium beam and a small cubic chamber which can be used for a 2D magneto-optical trap (MOT) to further collimate the atomic beam. The second part, with vacuum on the order of $10^{-9}$~mbar consists of a Zeeman slower and the science chamber. The design of the Zeeman slower used in both Sr1 and Sr2 systems is described in details in Ref. \cite{Bober10}. 
It has the capture velocity of 450~m/s and produces atoms slowed down to 30~m/s. The flux at the exit of the slower is  $3.5\times 10^{9}$~s${}^{-1}$ at the oven temperature of $460^{\circ}$C.

The science chamber in Sr1 is made of non-magnetic austenitic 316L stainless steel (Kimball Physics 8.0" Extended Spherical Octagon), while the chamber and its surroundings in Sr2 are made of fortal aluminium alloy and titanium (modified version of the chamber described in Ref. \cite{LeTargat07}).  All viewports flanges, nuts and bolts in scientific chamber of Sr2 are made of either aluminium or titanium. Carcasses of all coils are made of cooper. Non-magnetic environment is essential for stability of the clock since the ${}^{1}S_{0}$ -- ${}^{3}P_{0}$ transition in bosonic strontium is forbidden and a magnetic field is needed to induce non-zero dipole moment.

\subsection{Laser Cooling}

\begin{figure}[hb]
\centering
\includegraphics[width=0.9\columnwidth]{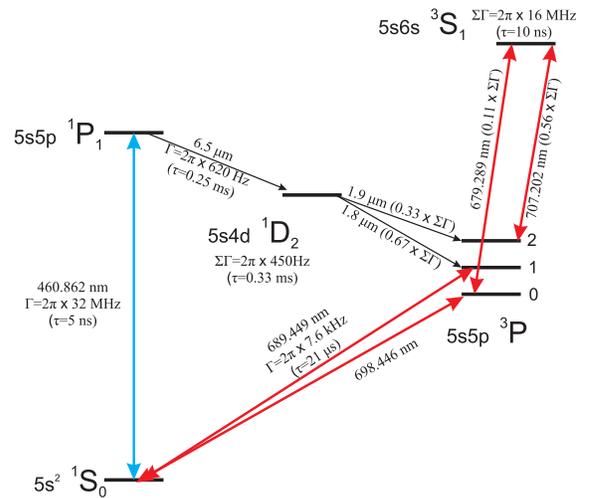}
\caption{(Color online) A simplified level structure of Sr atoms with transitions used in the experiment: first cooling stage -- 461~nm, second cooling stage -- 689~nm, clock
transition -- 698~nm, repumping -- 679 and 707~nm.
The natural lifetimes of the states  $\tau$ and the probabilities of the transitions $\Gamma$ are also indicated on the diagram. The inverse of the lifetime of the state is equal to the sum $\Sigma\Gamma$ of the probabilities of the transitions to  lower states. 
 \label{fig:lvl}}
\end{figure}

A simplified level structure of Sr atoms with transitions used in the experiment is presented in Fig. \ref{fig:lvl}. 
The cooling process of strontium atoms is performed in two stages: the first is the pre-cooling down to temperatures of a~few~mK, followed by ultimate laser cooling to temperatures below $10~\mu$K, low enough to capture atoms by the optical lattice trap. The first stage takes place in a blue MOT. For this cooling the strong allowed transition ${}^{1}S_{0}$ -- ${}^{1}P_{1}$ is used at 461~nm. 
Blue MOT beams are detuned 40~MHz  below the strontium
 ${}^{1}S_{0}$ -- ${}^{1}P_{1}$ transition and have 23~mm in diameter. To avoid the
situation where cooling 
is interrupted
by atoms relaxing 
to some other metastable states rather than to the ground state,
 two repumping lasers are used: 679 and 707~nm (${}^{3}P_{0}$ -- ${}^{3}S_{1}$ and ${}^{3}P_{2}$ -- ${}^{3}S_{1}$ transitions, respectively). 
In total, 6-8$\times 10^{8}$ atoms are loaded into the blue MOT and cooled down to 2-3~mK. The 461~nm lasers 
in two set-ups consist of one Extended Cavity Diode Lasers (ECDLs), emitting at 922 nm, amplified by two Tapered Amplifiers (TA) and frequency doubled
in the bow-tie cavities (modified Toptica TA-SHG pro systems). 
The blue light is locked to the  ${}^{1}S_{0}$ -- ${}^{1}P_{1}$ transition in bosonic ${}^{88}$Sr using the saturation spectroscopy in a hollow cathode lamp (Hamamatsu L233-38NB). 
The chain of the acousto-optic modulators (AOM) allows trapping either ${}^{88}$Sr or  ${}^{87}$Sr atoms in the blue MOTs.
 The frequencies of the repumping lasers are controlled by the wavelength meter (HighFinesse \AA{}ngstrom WS-6). The lasers are not locked to any other frequency reference.  A 10~kHz frequency modulation is applied to the repumping lasers with the amplitude of $\sim$500~MHz for ${}^{88}$Sr and $\sim$4~GHz for ${}^{87}$Sr to cover all  states to be repumped and to neglect frequency fluctuations in the free running ECDLs.
Modulation is applied to both the laser drive current and the laser PZT.

The second cooling stage takes place in a red MOT on the narrow ${}^{1}S_{0}$ -- ${}^{3}P_{1}$ transition at 689~nm. The natural width of this transition implies that the trapping laser has to be spectrally narrowed by a lock to a high-finesse optical cavity used as a short-term frequency reference (custom-made by Advanced Thin Films).
The design of the cavity and the 689~nm laser system is described in details in Ref. \cite{Lisak12}. 
In short, the Pound-Drever-Hall (PDH) locking scheme \cite{Drever83} is used to narrow the laser line width well below the width of the  ${}^{1}S_{0}$ -- ${}^{3}P_{1}$ transition.
 The cavity mirrors and the 100~mm spacer are made of ultra-low expansion glass. The mirrors are optically contacted to the spacer. The FSR of the cavity is 1.5~GHz and its finesse is $\mathcal{F} = 62800$. The cavity should be insensitive to vibrations, therefore a special design of the cavity is used with the horizontal cylindrical shape.
The  shape of the cavity is undercut to minimise impact of mechanical noises on the resonance frequencies \cite{Webster07}. 
  The cavity is placed on four viton spherules.
 The proper choice of the points where the supporting spherules are placed assures immunity to vibrations. 
 The cavity is isolated from the laboratory environment and enclosed in a vacuum chamber placed in another thermal enclosure. The temperature of the enclosure is actively stabilised with the accuracy of 10~mK. The passive vibration isolation platform (Minus K, BM-1) and a steel chamber lined with acoustic-damping foam (Novascan NanoCube) assure good mechanical isolation. 

The frequency of light from the ECDL (Toptica DL pro) stabilised to the cavity is digitally locked to the atomic ${}^{1}S_{0}$ -- ${}^{3}P_{1}$ transition by the acousto-optic frequency shifter. The transition is probed by the saturation spectroscopy technique in a separate vacuum chamber with an oven as a source of the atomic beam similar to the one in the main vacuum systems. 

The trapping lasers of the red MOTs in Sr1 and Sr2  are the Fabry-Perot diode lasers injection-locked to the light from ultra-stable 689 nm laser. The master-slave system amplifies the power of the red light and filters out any fluctuations of power of the injection laser.
These lasers (and all other slave lasers in the experiment) are placed in the mounts originally designed for distributed feedback diodes (Toptica ColdPack) which assures superior thermal stability. 

The red MOT beams have a diameter of~6 mm and are superimposed on the blue MOT beams. The magnetic field gradient is lowered from 0.55~T/m during the blue MOT phase to 0.03~T/m at the beginning of the red MOT phase.
Subsequently, the atomic cloud is compressed by linearly ramping the magnetic field up to 0.10~T/m. 
The natural width of the ${}^{1}S_{0}$ -- ${}^{3}P_{1}$ transition at 689~nm
 is much smaller than the Doppler width of this transition even at a temperature of a~few~mK (i.e., the
temperature of atoms in the blue MOT). 
To transfer as many atoms as possible from the blue MOT, in the first phase of the red MOT the narrow 
red laser beams have to be artificially broadened. 
A 20~kHz frequency modulation with the amplitude of 1.6~MHz is added by the AOM for that sake. 

In the case of ${}^{87}$Sr, a two-color red MOT is necessary, with the ${}^{1}S_{0}$(F = 9/2)  -- ${}^{3}P_{1}$ (F = 11/2) transition for the trapping and cooling process and the 
${}^{1}S_{0}$(F = 9/2)  -- ${}^{3}P_{1}$ (F = 9/2) transition (so-called ''stirring'' frequency\cite{Mukaiyama03}) for optical pumping of the ground states.

\subsection{Optical lattice}

In both Sr1 and Sr2 systems the optical lattices utilize an optical buildup
cavities with finesses  of around $\mathcal{F} =  120$.
In both realizations the mirrors of the cavity are mounted outside of the vacuum chamber. 
The TEM${}_{00}$ mode of the cavity at the wavelength of 813~nm has a waist of $w = 152~\mu$m and $w = 108~\mu$m in Sr1 and Sr2, respectively.  {The relatively large waists enable reduction of the density of captured atoms and hence the collisional shift of the clock line.}

The magic frequency for the  ${}^{87}$Sr and ${}^{88}$Sr isotopes is known to be 368~554~693(5)~MHz\cite{Westergaard11}  and 368~554.58(28)~GHz\cite{Akatsuka10}, respectively, which is close to the wavelength of 813~nm. The 813~nm light is generated by the ECDL - TA (Toptica TA 100).
The required level of 
control of the lattice-related effects implies that the frequency of the 813~nm laser has to be tuned with the accuracy of 1~MHz. The frequency of the 813 laser is locked to a Fabry-Perot transfer cavity. The cavity itself is referenced to the 
ultra-stable 689 nm (red MOT) laser 
stabilised to the  ${}^{1}S_{0}$ -- ${}^{3}P_{1}$ transition in 
  ${}^{88}$Sr.
 The frequency of the 813~nm light is further tuned by the AOMs. The light is transmitted and
spatially filtered through optical fibers to both Sr1 and Sr2 systems and injected into the cavities surrounding the atomic samples. The length of the cavities is stabilised to the 813~nm light by the PDH lock. The light coupled out from the cavity is observed by a fast photodiode. The photodiode signal is used by the PI controller and the AOMs for stabilisation of the power inside the cavity.  {The wavemeter used in the experiment (HighFinesse WS6/200) allows one to pre-tune the frequency of the 813~nm laser  with accuracy of 200 MHz. More precise tuning has to be done by observing the light-shift of the clock line.}

\section{Ultra-stable laser system}

\begin{figure}[hb]
\centering
\includegraphics[width=\columnwidth]{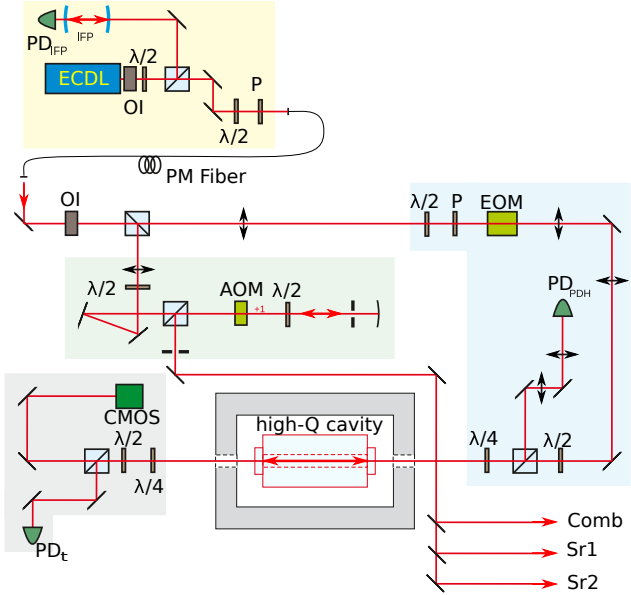}
\caption{(Color online) The optical scheme of the 698~nm laser. ECDL is the Extended Cavity Diode Laser, OI is the Optical Isolator, EOM is the electro-optic phase modulator, AOM is the acousto-optic modulator, IFP is the scanning Fabry-Perot interferometer, FR 	is the Faraday rotator, P is the Glan-Taylor polariser, PD${}_{t}$, PD${}_{IFP}$, PD${}_{PDH}$
 are the photodiodes, CMOS is the CMOS camera, $\lambda/2$ is the half-wavelength plate, $\lambda/4$ is the quarter-wavelength plate
 \label{fig:698}}
\end{figure}

An ECDL laser (Toptica DL pro) locked to the TEM${}_{00}$ mode of the high-Q cavity is the short-time frequency reference of the optical standards. 
The optical scheme of the 698~nm laser system is depicted in Fig. \ref{fig:698}. The following subsections describe areas marked by different colors of the background.

\subsection{Ultra-stable laser}

The heart of the system is the high-Q cavity (Stable Laser Systems, type ATF 6010-4) made of the ultra-low expansion glass (ULE). The FSR of the cavity is 1.5~GHz and its finesse is $\mathcal{F} = 300000$. The spacer (100~mm long) is made of the ULE, while the mirrors are made of  silica. One of the mirrors is flat, the other one is concave with the radius $R=500$~mm. Similarly to the 689~nm cavity, the cylindrical shape of the cavity is undercut to minimise impact of mechanical noises on the resonance frequencies. The cavity is placed on four viton spherules on the zerodur support  {inside the vacuum chamber.}

\begin{figure}[htb]
\centering
\includegraphics[width=0.6\columnwidth]{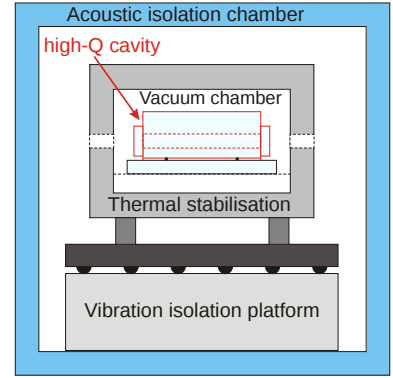}
\caption{(Color online) The scheme of the thermal, mechanical and acoustical isolation of the high-Q cavity.
 \label{fig:cavity}}
\end{figure}

   {The vacuum in the cavity chamber is on the order of the $10^{-6}$ Torr.} The thermal shielding of the chamber provides passive isolation (thermal conductance of 0.5 W/K). The active stabilisation of the temperature (a PID controller and two heaters) has the accuracy of 2~mK. The chamber and the optical set-up which is needed for the PDH lock is placed on the vibration isolation platform (Minus K, BM-1). The damping efficiency of the platform is in the range between -20~dB and -60~dB for the frequencies of 2~Hz and 100~Hz, respectively. The platform and the chamber are enclosed in the acoustic isolation chamber (Novascan NanoCube) providing isolation on the level of -40~dB. The  scheme of the thermal, mechanical and acoustical isolation of the high-Q cavity is sketched in Fig. \ref{fig:cavity}.  {Finite thermal stability of our high-Q cavity and small fluctuations of the vacuum pressure in the cavity chamber result in the overall instability of our cavity measured relative to one of the optical standards equals to $3\times 10^{-15}$ after 10~s, $7\times 10^{-15}$ after 100~s and $1.5\times 10^{-14}$ after 200~s. For times longer than 200~s the fluctuations of the cavity start to average.}

The yellow area in  Fig. \ref{fig:698} marks the section including the ECDL laser (the spectral width of 50~kHz when free-running) and a scanning confocal Fabry-Perot interferometer (IFP) with a photodiode (PD${}_{IFP}$) monitoring the single-mode operation of the laser. This section is placed on  {a low-profile} vibration isolation platform (Herzan Onyx-6M) and enclosed in a chamber made of  PVC foam.  {This reduces the acoustic noises from the outside environment and helps to keep the ECDL laser frequency jumps within the range of the PDH lock to the cavity.} The light from the ECDL laser is transferred into the cavity chamber by a PM fiber.

The blue area in  Fig. \ref{fig:698} marks the optical section needed for locking of the  ECDL laser to the cavity mode by the PDH method. The electro-optic phase modulator (EOM) 
working at 20 MHz adds sidebands to the laser frequency.
The incident light has the power below $10~\mu$W. The fast silicone avalanche photodiode (PD${}_{PDH}$) (Hamamatsu C5658) observes the signal of the beatnote between the light incident on and reflected from the cavity. The beatnote signal is demodulated with the mixer. The resulting error signal is used by the fast PID regulator (Toptica FALC) to stabilise the frequency and to narrow the spectral width of the ECDL laser by controlling the laser diode current and the length of the ECDL internal resonator.

The gray area in  Fig. \ref{fig:698} contains a CMOS camera and a silicone photodiode (PD${}_{t}$) which are used for monitoring the power and quality of the  TEM${}_{00}$ mode of the cavity.

In the green area in  Fig. \ref{fig:698} an AOM in the double-pass set-up is used for real-time correction of the slow drift (typically less than 0.1~Hz/s) of the frequency of the cavity modes.	

\subsection{Doppler cancellation of the fiber-link noises}

\begin{figure}[htb]
\centering
\includegraphics[width=0.9\columnwidth]{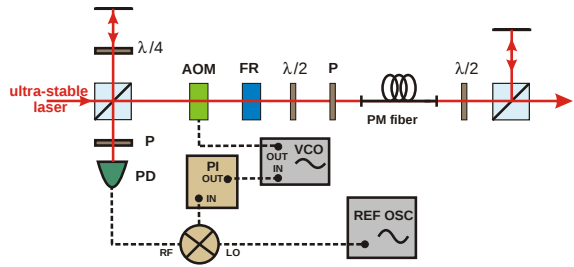}
\caption{(Color online) The scheme of the system for Doppler cancellation of the fiber-link noises. AOM is the acousto-optic modulator, FR 	is the Faraday rotator, P is the Glan-Taylor polariser, PD is  the photodiode, VCO is the Voltage Controlled Oscillator, REF OSC is the reference oscillator, $\lambda/2$ is the half-wavelength plate, $\lambda/4$ is the quarter-wavelength plate.
 \label{fig:noise}}
\end{figure}

The light from the ultra-stable laser is transferred to the Sr1 and Sr2 standards and to the optical frequency comb through fibers. 
Each fiber has a system 
of active Doppler cancellation of the fiber-link noises to assure the transfer of stable optical frequencies
\cite{Ma94}. 
The complete scheme of the system for the 
Doppler cancellation is presented in Fig. \ref{fig:noise}.	The stable frequency of the ultra-stable laser is shifted by $\omega_s$ in the acousto-optical modulator (AOM). The light is subsequently transmitted through a polarisation-maintaining fiber. Small part of the light is reflected, transmitted back through the fiber and AOM and superimposed with a portion of the original beam. The photodiode (PD) measures the beatnote  of these two beams. The signal from the photodiode (RF)
is demodulated in the mixer with the local oscillator signal (LO) from the stable reference oscillator (REF OSC) with the frequency of 2$\omega_s$. The output of the mixer is the error signal proportional to the cosine of the phase difference between the original beam and the round-trip beam which passed twice through the fiber and AOM. The error signal through the PI integrator controls the voltage controlled oscillator (VCO). The VCO generates the RF signal for the AOM and with the feedback control PLL loop 
 cancels the phase noises in the optical signal transmitted through the fiber.  

\subsection{Clock beams in Sr1 and Sr2 standards}

In both Sr1 and Sr2 systems the Fabry-Perot diode lasers are injection-locked to the light from ultra-stable 698~nm laser. The master-slave system  filters out any power fluctuations of the injection laser. The residual amount of spectrally broadband Amplified Spontaneous Emission (ASE) of an injection-locked diode is low, since the locked diode operate well above threshold with a strongly saturated gain. Nevertheless, the narrowband interferometric filters with FWHM of the transmition peak below  0.4~nm (custom-made by ATFilms) are placed in the beam to ensure that only the ultra-narrow light will pass to the atoms. The beam is passing the AOM of the digital lock and is  {injected} to the cavity of the optical lattice such that the clock beam is exactly superimposed with the lattice and its waist is much bigger than the size of the sample of atoms.

\subsection{Optical frequency comb}

The frequency of the clock transition in our experimental  {set-up} is measured with GPS-disciplined Rb frequency standard by the use of Er:fiber polarization mode-locked optical frequency comb (Menlo FC 1500-250-WG). Its repetition rate equals to 250 MHz and can be changed in the wide range of 2 MHz.
To reach the Sr clock transition wavelength, the fundamental output of the laser around 1550~nm is spectrally shifted in the Er:fiber amplifier to $\sim$1396~nm and subsequently frequency doubled in a periodically-polled lithium niobate  (PPLN) crystal, which results in approximately 10~mW of power in 6~nm wide spectrum around 698~nm.
The optical frequency comb is fully stabilized to the reference radio frequency source. Additionally, the high bandwidth ($>500$~kHz) intra-cavity electro-optic phase modulator (EOM) can be used, which allows phase locking it to our clock laser as the optical frequency reference.
This enables coherent transfer of optical phase between our clock laser and the near-infrared telecommunication range at $10^{-16}$ short-term stability level \cite{Hagemann13} for future transfer of ultra-stable optical frequencies through the fiber networks or referencing high-resolution spectroscopy in that range \cite{Foltynowicz11,Wojtewicz13,Truong13,Ehlers12,Amodio14}.

A new possibility of the absolute frequency measurement of the clocks is opened by installation of a long distance stabilized fiber optic link connecting our lab to UTC(AOS) and UTC(PL) via the OPTIME network \cite{Sliwczynski13}. In this scheme the Er:fiber optical frequency comb will be locked to the RF frequency disseminated via the fiber.

\section{Results}

To compare two  lattice clocks both standards were tuned to the ${}^{1}S_{0}$ -- ${}^{3}P_{0}$ transition in bosonic ${}^{88}$Sr.
Comparing two clocks using the same atomic species assures that no systematic effects have been overlooked in their individual accuracy budget, since the measured frequencies should be the same within the accuracy budgets of both standards. 

\subsection{The time sequence and interrogation}

\begin{figure}[hb]
\centering
\includegraphics[width=\columnwidth]{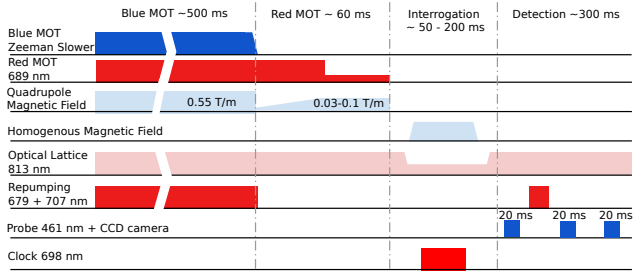}
\caption{(Color online) The timing scheme of one cycle. \label{fig:time}}
\end{figure}

Lattice clock operates in precisely defined cycles. In each cycle atoms are prepared before the interrogation, then the ultra-stable laser interrogates the clock transition and finally the populations in the ground and excited states are measured, which provides information on the transition probability. The time sequence of one cycle is presented in Fig. \ref{fig:time}. After loading the atoms into the optical trap, the atoms are probed  by a single $\pi$-pulse. 
The electric-dipole ${}^{1}S_{0}$ -- ${}^{3}P_{0}$ transition in bosonic ${}^{88}$Sr is forbidden, hence for inducing of the non-zero dipole moment the MOT coils are switched from the anti-Helmholtz to the Helmholtz configuration and a homogeneous magnetic field (up to 26~mT) is applied to the atoms \cite{Taichenachev06}.

The absolute probability that the atoms changed their state from ${}^{1}S_{0}$ to ${}^{3}P_{0}$ is calculated from the measurements of populations $N_g$ and $N_e$ of the ground and excited state, respectively, and is equal to $P=N_e/(N_g+N_e)$. 
After the interrogation by the clock light, the population $N_g$ is measured by fluorescence imaging with strong 461 nm transition. 
All measurements are done by recording the fluorescence images by the electron multiplying CCD camera (Hamamatsu C9100-02).
The pulse of the blue light blows  the ground-state atoms out of the lattice trap while the excited atoms are shelved in the dark state for that transition. Sequentially, the 679 and 707~nm repumping lasers  pump the excited atoms back to the ground state and  their number, $N_e$ is measured by the second fluorescence imaging at the 461 nm transition. A third photo which gives the background is  made without any atoms in the trap.

\subsection{Resolved Sideband Spectroscopy}

\begin{figure}[ht]
\centering
\includegraphics[width=\columnwidth]{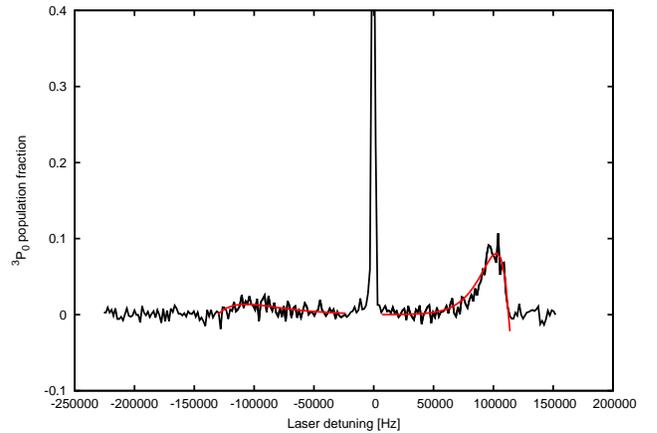}
\caption{(Color online) Spectroscopy of the longitudal sidebands of the clock transition. An analytic formula \cite{Blatt09} (red) is fitted to the data (black).\label{fig:side}}
\end{figure}

In principle, the atoms which have been loaded into the optical trap after cooling in the red MOT have the temperature much lower than the depth of the trap. Therefore, they should mostly occupy  lower motional states of the potential. This can be verified by the spectroscopy of the longitudinal sidebands of the clock transition \cite{Blatt09}. The clock line ${}^{1}S_{0}$ -- ${}^{3}P_{0}$, where  the motional state $\vert n \rangle$ is conserved is  accompanied by a red and blue sidebands  where
the motional state decreases ($\vert n \rangle \rightarrow \vert n-1 \rangle$) and increases ($\vert n \rangle \rightarrow \vert n+1 \rangle$) by 1, respectively. The atoms which are in the motional ground state  $\vert n = 0 \rangle$ cannot contribute to the red sideband transition, hence  the longitudinal temperature can be measured from the relative height of the two sidebands. The width of the sidebands provides the information of the transverse temperature in the lattice trap. Additionally, the positions of the sidebands can be used for calibration of the trap depth. An example of the sideband spectrum is presented in Fig. \ref{fig:side}. The measured lattice depth is $U_0 = 338 E_r \approx 56$~$\mu$K, where $E_r=h^2/(2m\lambda^2)$ is the lattice recoil energy. The fitted expressions for the probability of the transition (see Appendix in the Ref. \cite{Blatt09}) resulted in  the longitudinal and transverse temperatures  equal to $T_z=2.5$~$\mu$K and $T_r=30$~$\mu$K, respectively.

\subsection{Spectroscopy of the ${}^{1}S_{0}$ -- ${}^{3}P_{0}$ transition, stability of the strontium optical lattice clocks }

\begin{figure}[ht]
\centering
\includegraphics[width=\columnwidth]{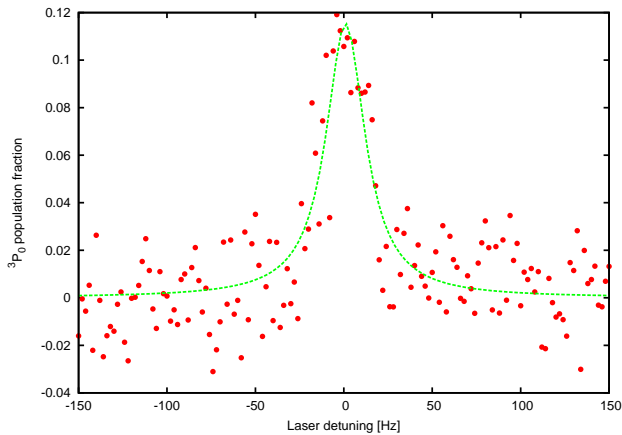}
\caption{(Color online) Spectroscopy of the ${}^{1}S_{0}$ -- ${}^{3}P_{0}$ transition. A Lorentz function is fitted to the measured data. The measured  linewidth (FWHM) is equal to 27(1)~Hz. \label{fig:line}}
\end{figure}

An example of the spectroscopy of the clock transition is presented in Fig. \ref{fig:line}. The atoms were interrogated by a 50~ms  pulse with the intensity of 400~mW/cm${}^2$ in the magnetic field of 0.67~mT. A Lorentz function is fitted to the measured data. The measured  linewidth (FWHM) is equal to 27(1)~Hz and is mostly limited by the pulse duration.

\begin{figure}[ht]
\centering
\includegraphics[width=\columnwidth]{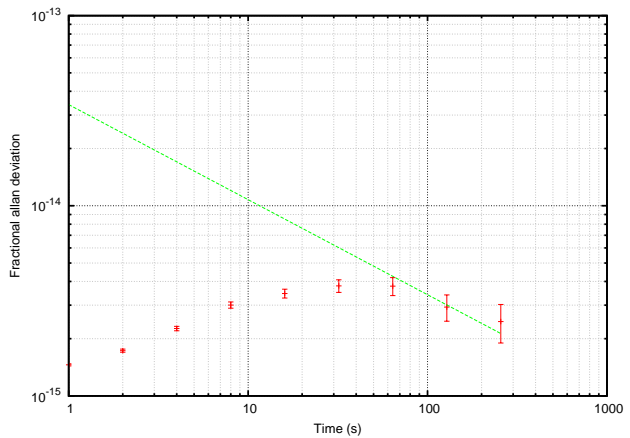}
\caption{(Color online) The measured frequency stability (frequency difference between two standards) in fractional units represented by the Allan standard deviation.
The dashed (green) line shows the asymptotic stability of $\sigma_y(\tau) = 3.41(27) \times 10^{-14}/\sqrt{\tau}$.

\label{fig:allan}}
\end{figure}

To probe independently each of the standards, the clock laser beam is split into two parts and  sent to the two standards by separate acousto-optic frequency shifters.
The frequencies of the beams are locked to the clock transition with a digital integrator by comparing the absorption signal at the opposite half-width points. Sequentially, the correction is applied to the driving frequency of AOM. 
 The difference between the corrections in both standards gives the momentary frequency difference between two clocks. 
The measured frequency stability in fractional units represented by the Allan standard deviation is presented in Fig. \ref{fig:allan}.
For the average times $\tau$ greater than 60~s the Allan deviation  decreased with $\sigma_y(\tau) = 3.41(27) \times 10^{-14}/\sqrt{\tau}$. This value is close to the $\sigma_y(\tau) = 2.3 \times 10^{-14}/\sqrt{\tau}$ obtained by Katori {\em et al.} \cite{Akatsuka08,Akatsuka10} for asynchronous operation of two clocks,  a 3D lattice clock with bosonic ${}^{88}$Sr and a 1D lattice clock with fermionic ${}^{87}$Sr. The synchronous excitation, which  in our system will be implemented in the near future, allowed the group of Katori to improve the stability of their bosonic clock to $\sigma_y(\tau) = 4 \times 10^{-16}/\sqrt{\tau}$ \cite{Takamoto11}.

\subsection{Preliminary uncertainty budget and relative frequency difference measured between two standards \label{sec:shifts}}

 {We have evaluated the main contributions to the frequency 
shifts in both standards and compared them in Table \ref{tab:acc}.}

 {Two most important contributions to the budget are the quadratic Zeeman shift and the light shift from the 698 clock laser. 
Both shifts were evaluated by making two sets of four simultaneous (interlaced) locks to the atomic line with four different values of the magnetic field or four different intensities of the clock laser, respectively. The nonlinear Zeeman shift was calibrated by fitting a quadratic function to the result of the first set of measurements, while the 698 nm light shift was calibrated by fitting a linear function to the result of the second set of data.}

 {The scalar light shift from the 813 lattice laser was evaluated by calculating the shift corresponding to detuning of  the lattice light frequency from the magic value \cite{Akatsuka08}. We used the wavemeter accuracy, i.e. 200 ~MHz, with the coefficient estimated from data in  Ref. \cite{Brusch06}.}

 {The shifts induced by the blackbody radiation (BBR)  can be described as static shifts with a small dynamic correction \cite{middelmann1}. The static contribution is proportional to the differential static polarisability of the two clock states $\Delta\alpha = 4,07873(11) \times 10^{-39}$~ Cm${}^2$/V \cite{middelmann2,marianna} and the mean square value of the electric field at temperature $T$ (scales with $T$ like $T^4$). The dynamic contribution of the $5s^2$ $^1S_0$ ground state is calculated for the $5s5p$ $^3P_1$ and $5s5p$ $^1P_1$ transitions. To calculate the dynamic contribution of the $5s5p$ $^3P_0$ excited state four transitions are taken into account: $5s4d$ $^3D_1$, $5s6s$ $^3S_1$, $5p^2$ $^3P_1$ and $5s5d$ $^3D_1$ (see the  discussion in Ref. \cite{marianna,porsev2}). The dynamic correction is calculated following Ref. \cite{middelmann2} and the transition data are taken from Ref. \cite{marianna,chem}. }

 {The temperature of a few crucial points of the vacuum system is monitored during the experiment cycle by MC65F103A thermistors. The acquired data and an accurate model of the vacuum system is used to  set a finite elements stationary thermal simulation to obtain the temperature distribution of the system. 
This simulation is used to calculate the BBR experienced by the atoms.
The uncertainty of the shift is evaluated from calculations of the BBR for maximum and minimum temperatures measured in the experiment.}

 {
The value of the collisional shift is evaluated by estimating the number of atoms per lattice site. For known lattice waists the atomic density is estimated by measuring the fluorescence intensity of the lattice-trapped atoms with a calibrated photodiode and number of occupied sites with a CCD camera.
The frequency shift is calculated following Ref.  \cite{Lisdat09}. The uncertainty of this shift is measured by making the set of four simultaneous (interlaced) locks to the atomic line with four different numbers of atoms loaded into the lattice. Since varying the atom number we do not see any shift we take measured data as the upper limit  for possible collisional shift.
}

 {
The last evaluated uncertainty represents the finite resolution of the DDSs driving the AOMs  in the frequency chain of the clock lasers.}

 {
The relative frequency difference measured between two standards, taking into account the accuracy budget presented above, is equal to 19(42)~Hz.}

\begin{table}[h!]

\caption{Accuracy budget for typical experimental conditions. All numbers are in Hz.}

\label{tab:acc}
\centering

\begin{tabular}{ccc}
           \br
            Effects & \multicolumn{2}{c}{Shift(Uncert.)}  \\ 

             ~& Sr1 & Sr2\\
            \br
            Quadratic Zeeman & -132(37)& -126(19) \\ 
            \mr 
            Probe light & -33.3(1.5) & -8.1(1.6) \\ 
            \mr 
            Lattice light & 0.0(1.2) & 0.0(1.2)\\
            \mr
            Blackbody radiation & -2.210(0.075) & -2.405(0.075) \\ 
            \mr 
            Collisions & 0.1(1.0) & 0.1(1.0) \\ 
            \mr 
            DDS \& electronics & 0.00(0.23) & 0.00(0.12) \\ 
            \br
            {\bf Total:} & {\bf -167(37)} & {\bf -136(19)}\\
            \br
            \end{tabular} 
\end{table}

\subsection{Absolute frequency measurement of the ${}^{1}S_{0}$ -- ${}^{3}P_{0}$ transition}

 {
The absolute frequency of the clock transition in our experimental set-up can be measured  by the use of the Er:fiber  optical frequency comb. At this stage of the experiment the optical frequency comb is locked to the GPS-disciplined standard which limits its accuracy on the level of $10^{-12}$ \cite{Morzynski13}. Within this precision the measured absolute frequency of the clock transition in Sr1 is equal to 429~228~066~418~300(580)~Hz. This value agrees within its uncertainty with the value recommended by BIPM \cite{BIPM88}.  We believe that after connecting to the OPTIME network \cite{Sliwczynski13} we will be able to calibrate the absolute frequency directly to the hydrogen maser and improve the precision below $10^{-15}$.
}

\section{Conclusion}

We presented preliminary results of spectroscopy of the clock line and frequency stability of the two  strontium optical lattice clocks. 
These standards are interrogated by a shared ultra-narrow laser pre-stabilised  to  a high-Q optical cavity. The frequency of the clock transition in our experimental setup can be compared with the GPS-disciplined Rb frequency standard by the use of Er:fiber polarization mode-locked optical frequency comb. 

 {
In the current state of the experiment, the fluctuations of the relatively high magnetic field, corresponding to the clock transition line-width of few tens of Hz, limit stability of our clocks to about $2\times 10^{-15}$. The accuracy of the Sr2 system, according to our preliminary uncertainty budget, is better than $5\times 10^{-14}$. Given the BIPM limits are above $1\times 10^{-14}$ for the bosonic clocks, such  stability and accuracy are sufficient for our present goals.
}

The reliable operation and good stability of the setup, allows one to expect good performance of the system after next steps of improvements.
In the near future the  absolute frequency of the clocks will be verified by the use of the optical frequency comb and a long distance stabilized fiber optic link with UTC(AOS) and UTC(PL) via the OPTIME network \cite{Sliwczynski13}.

\section*{Acknowledgement}

The authors would like to thank Dr J\'erome L\^odewyck and Dr Rodolphe Le Targat for valuable discussions and help in designing the optical lattice standards.
This work has been performed in the National Laboratory FAMO in Toru\'n and supported by the
subsidy of  
the
Ministry of Science and Higher  { Education.}
 {
Individual contributors were partially supported by
the Polish National Science Centre Projects No. 2012/07/B/ST2/00235, No. UMO-2013/11/D/ST2/02663, No. 2012/07/B/ST2/00251, No. 2012/05/D/ST2/01914 and by  the Foundation for Polish Science  Projects Start, Homing Plus and the TEAM Project co-financed by the EU within the European Regional Development Fund.
}


\section*{References}
\bibliographystyle{unsrt}

\begin{thebibliography}{10}

\bibitem{Ido03} Ido T, and Katori H 2003 Recoil-Free Spectroscopy of Neutral Sr Atoms in the Lamb-Dicke Regime {\em Phys. Rev. Lett.} {\bf 91} 053001-4


\bibitem{Rosenband08}  Rosenband T,  Hume DB,  Schmidt PO,  Chou CW,  Brusch A,  Lorini L,  Oskay WH,  Drullinger RE,  Fortier TM,  Stalnaker JE,  Diddams SA,  Swann WC, Newbury NR,  Itano WM,  Wineland DJ,  and  Bergquist JC 2008 Frequency Ratio of Al${}^+$ and Hg${}^+$ Single-Ion Optical Clocks, Metrology at the 17th Decimal Place {\em Science} {\bf 319} 1808



\bibitem{Ye14} Bloom BJ, Nicholson TL, Williams JR, Campbell SL, Bishof M, Zhang X, Zhang W, Bromley SL and Ye J 2014
An optical lattice clock with accuracy and stability at the $10^{−18}$ level {\em Nature} {\bf 506} 71-75

\bibitem{LeTargat13}  Le Targat R, Lorini L, Le Coq Y, Zawada M, Gu\'ena J,	Abgrall M, Gurov M,	Rosenbusch P, Rovera DG, Nag\'orny B,
Gartman R, Westergaard PG, Tobar ME, Lours M, Santarelli G, Clairon A, Bize S, Laurent P, Lemonde P and Lodewyck J 2013
Experimental realization of an optical second with strontium lattice clocks {\em Nature Commun.} {\bf 4} 2109

\bibitem{Hinkley13} Hinkley N, Sherman JA, Phillips NB, Schioppo M, Lemke ND, Beloy K, Pizzocaro M, Oates CW and
Ludlow AD 2013 An atomic clock with $10^{-18}$ instability {\em Science} {\bf 341} 1215-8

\bibitem{Falke14} Falke S, Lemke N, Grebing C, Lipphardt B, Weyers S, Gerginov V, Huntemann N, Hagemann C, Al-Masoudi A and H\"afner S 2014
A strontium lattice clock with $3\times10^{-17}$ inaccuracy and its frequency 
{\em New J. Phys.} {\bf 16} 073023

\bibitem{Ushijima14} Ushijima I, Takamoto M, Das M, Ohkubo T and  Katori H 2015
 Cryogenic optical lattice clocks {\em Nature Photon.} {\bf 9} 185-189

\bibitem{BIPM87} Bureau International des Poids et Mesures (BIPM) 2013 {\em Recommended Values Of Standard Frequencies
For Applications Including The Practical Realization
Of The Metre And Secondary Representations Of The
Definition Of The Second, Strontium 87 Atom ($f\approx~429$ THz)}, (BIPM, S\`evres, France)
\bibitem{BIPM88} Bureau International des Poids et Mesures (BIPM) 2009 {\em Recommended Values Of Standard Frequencies
For Applications Including The Practical Realization
Of The Metre And Secondary Representations Of The
Definition Of The Second, Strontium 88 Atom ($f\approx~429$ THz)}, (BIPM, S\`evres, France)


\bibitem {Peik04} Peik E, Lipphardt B, Schnatz H,  Schneider T,  Tamm Chr 
and  Karshenboim SG 2004  Limit on
the present temporal variation of the fine structure constant
{\em Phys. Rev. Lett.}
{\bf 93} 170801-1-4,


\bibitem{Fortier07}  Fortier TM, Ashby N, Bergquist JC,  Delaney MJ,  Diddams SA,  Heavner TP, Hollberg L,
 Itano WM,  Jefferts SR,  Kim K,  Levi F,  Lorini L,  Oskay WH, 
  Parker TE,  Shirley J and  Stalnaker JE 2007
   Precision atomic spectroscopy for improved limits on variation of the fine structure constant and local position invariance
{\em Phys. Rev. Lett.}
{\bf 98} 070801

\bibitem{Blatt08} Blatt S,  Ludlow AD,  Campbell GK,  Thomsen JW,  Zelevinsky T,  Boyd MM,  Ye J,
 Baillard X,  Fouch\'e M,  Le Targat R,  Brusch A,  Lemonde P,  
 Takamoto M,  Hong F-L,  Katori H and  Flambaum VV 2008 New limits on coupling of fundamental constants to gravity using
${}^{87}Sr$ optical lattice clocks
{\em Phys. Rev. Lett.}
{\bf 100} 140801-1-4

\bibitem{Peik10}  Peik E 2010 Fundamental constants and units and the search for temporal variations
{\em Nuclear Physics B (Proc. Suppl.)}
{\bf 203-204} 18-32

\bibitem{Bjerhammar85} Bjerhammar A 1985 On a relativistic geodesy
{\em Bulletin G\'eod\'esique}
{\bf 59} 207-220

\bibitem{Delva13} Delva P and  Lodewyck J 2013 Atomic clocks: new prospects in metrology and geodesy 
{\em Acta Futura} {\bf 7} 67-78


\bibitem{Schiller09} Schiller S, Tino G,  Gill P,  Salomon C,  Sterr U,  Peik E,  Nevsky A,  G\"orlitz A,  Svehla D,
Ferrari G, Poli N, Lusanna L, Klein H, Margolis H,  Lemonde P, 
Laurent P,  Santarelli G,
 Clairon A,  Ertmer W,  Rasel E,  M\"uller J,  Iorio L,  L\"ammerzahl C,  Dittus H,  Gill E,
 Rothacher M,  Flechner F,  Schreiber U,  Flambaum V,  Ni W-T,
 Liu L,  Chen X,
 Chen J,  Gao K,  Cacciapuoti L,  Holzwarth R,  He\ss{} M and  Sch\"afer W 2009
 Einstein gravity explorer -- a medium-class fundamental physics mission
{\em Experimental Astronomy}
{\bf 23} 573-610

\bibitem{Chou10}  Chou CW,  Hume DB,  Rosenband T and  Wineland DJ 2010
 Optical clocks and relativity
{\em Science}
{\bf 329} 1630 - 1633


\bibitem{Derevianko14}  Derevianko A and Pospelov M 2014 Hunting for topological dark matter with atomic clocks {\em Nat. Phys.} {\bf 10} 933-936


\bibitem{Dicke53}  Dicke RH 1953 The Effect of Collisions upon the Doppler Width of Spectral Lines  {\em Phys. Rev.} {\bf 89} 472



\bibitem{Takamoto05} Takamoto M, Hong F-L, Higashi R and Katori H 2005 An optical lattice clock {\em Nature} {\bf 435}
 321-324

\bibitem{Ye08} Ye J, Kimble HJ and Katori H 2008 Quantum State Engineering and Precision Metrology Using State-Insensitive Light Traps {\em Science} {\bf 320} 5884

\bibitem{Lemonde05}  Lemonde P and Wolf P Optical lattice clock with atoms confined in a shallow trap 2005 {\em Phys. Rev. A}  {\bf 72} 033409


\bibitem{Bober10}  Bober M, Zachorowski J and Gawlik W 2010 Designing Zeeman slower for strontium atoms - towards optical atomic clock {\em Opt. Appl.} {\bf 40} 547-555

\bibitem{LeTargat07}  Le Targat R 2007 Horloge \`a r\'eseau optique au Strontium: une 2\'eme g\'en\'eration d'horloges \`a atomes froids {\em PhD Thesis} https://hal.archives-ouvertes.fr/pastel-00553253v1


\bibitem{Lisak12}  Lisak D, Cygan A, Bielska K, Piwinski M, Ozimek F, Ido T, Trawinski RS and Ciurylo R 2012 Ultra-Narrow Laser for Optical Frequency Reference {\em Acta Phys. Pol. A} {\bf 121} 614-621

\bibitem{Drever83}  Drever RWP,  Hall JL,  Kowalski FV,  Hough J,  Ford GM,  Munley AJ and Ward H 1983 Laser phase and frequency stabilization using an optical resonator {\em Appl. Phys. B} {\bf 31} 97-105

\bibitem{Webster07} Webster SA, Oxborrow M and Gill P 2007 Vibration insensitive optical cavity {\em Phys. Rev. A} {\bf 75} 011801(R)

\bibitem{Mukaiyama03} Mukaiyama T, Katori H, Ido T, Li Y and Kuwata-Gonokami M 2003 Recoil-Limited Laser Cooling of ${}^{87}$Sr Atoms near the Fermi Temperature {\em Phys.
Rev. Lett.} {\bf 90} 113002

\bibitem{Westergaard11} Westergaard PG, Lodewyck J, Lorini L, Lecallier A, Burt EA, Zawada M, Millo J and Lemonde P 2011 Lattice-Induced Frequency Shifts in Sr Optical Lattice Clocks at the $10^{-17}$ Level {\em Phys. Rev. Lett.} {\bf 106} 210801

\bibitem{Akatsuka10} Akatsuka T, Takamoto M and Katori H 2010 Three-dimensional optical lattice clock with bosonic ${}^{88}$ Sr atoms {\em Phys. Rev. A} {\bf 81} 023402

\bibitem{Ma94}  Ma LS,  Jungner P,Ye J, and Hall JH 1994 Delivering the same optical frequency at two places: introduced by an optical fiber or other time-varying path accurate cancellation of phase noise {\em Opt. Lett.} {\bf 19} 1777-1779

\bibitem{Hagemann13} Hagemann C, Grebing C, Kessler T, Falke S, Lemke N,  Lisdat C,  Schnatz H,  Riehle F and Sterr U 2013 Providing $10^{-16}$ Short-Term Stability of a 1.5-$\mu$m Laser to Optical Clocks {\em  IEEE Trans.  Instrum.  Meas.} {\bf 62} 1556-1562

\bibitem{Foltynowicz11}  Foltynowicz A,  Ban T,  Mas\l{}owski P,  Adler F and Ye J 2011 Quantum-Noise-Limited Optical Frequency Comb Spectroscopy {\em Phys. Rev. Lett.} {\bf 107} 233002

\bibitem{Wojtewicz13}  W\'ojtewicz S,  Stec K,  Mas\l{}owski P,  Cygan A,  Lisak D, . Trawi\'nski RS,  Ciury\l{}o R 2013 Low pressure line-shape study of self-broadened CO transitions in the $\left(3\leftarrow 0\right)$ band {\em J. Quant. Spectrosc. Radiat. Transf.} {\bf 130} 191-200

\bibitem{Truong13}   Truong G-W,  Douglass KO, Maxwell,	SE, van Zee RD, Plusquellic DF,  Hodges	JT and  Long DA 2013 Frequency-agile, rapid scanning spectroscopy {\em Nat.Photon.} {\bf 7} 532-534


\bibitem{Ehlers12} Ehlers P,  Silander I, Wang J and  Axner O 2012 Fiber-laser-based noise-immune cavity-enhanced optical heterodyne molecular spectrometry instrumentation for Doppler-broadened detection in the 10${}^{-12}$  cm${}^{-1}$ Hz${}^{-1/2}$ region {\em JOSA B} {\bf 29} 1305-1315

\bibitem{Amodio14} Amodio P, Moretti L,  Castrillo A and  Gianfrani L 2014 Line-narrowing effects in the near-infrared spectrum of water and precision determination of spectroscopic parameters {\em J. Chem. Phys.} {\bf 140} 044310 


\bibitem{Sliwczynski13}  \'Sliwczyński \L{}, Krehlik P,  Czubla A,  Buczek \L{} and Lipi\'nski M 2013 Dissemination of time and RF frequency via a stabilized fibre optic link over a distance of 420 km {\em Metrologia} {\bf 50} 133-145



\bibitem{Taichenachev06} Taichenachev AV, Yudin VI, Oates CW, Hoyt CW, Barber ZW and Hollberg L 2006 Magnetic field-induced spectroscopy of forbidden optical transitions with application to lattice-based optical atomic clocks {\em Phys. Rev. Lett.} {\bf 96} 083001

\bibitem{Blatt09}  Blatt S,  Thomsen KW, Campbell GK,  Ludlow AD,  Swallows MD,
 Martin MJ,  Boyd MM and  Ye J 2009 Rabi spectroscopy and excitation inhomogeneity in a one-dimensional optical lattice clock {\em Phys. Rev. A} {\bf 80} 052703

\bibitem{Akatsuka08} Akatsuka T, Takamoto M and Katori H 2008 Optical lattice clocks with non-interacting bosons and fermions 2008 {\em Nature Phys.} {\bf 4}  954-959 


\bibitem{Takamoto11} Takamoto M, Tetsushi T and Katori H 2011 Frequency comparison of optical lattice clocks beyond the Dick limit {\em Nature Photon.} {\bf 5} 288-292

\bibitem{Brusch06}  Brusch A, Le Targat R, Baillard X, Fouch\'e M and Lemonde P 2006 Hyperpolarizability Effects in a Sr Optical Lattice Clock {\em Phys. Rev. Lett.} {\bf 96} 103003 


\bibitem{middelmann1} Middelmann T,  Lisdat C,  Falke S,  Vellore Winfred JSR,  Riehle F and  Sterr U 2011 Tackling the blackbody shift in a strontium optical lattice clock {\em IEEE Trans. Instrum. Meas.} \textbf{60}, 2550 

\bibitem{middelmann2} Middelmann T, Falke S,  Lisdat C and  Sterr U 2012 High Accuracy Correction of Blackbody Radiation Shift in an Optical Lattice Clock {\em Phys. Rev. Lett.} \textbf{109}, 263004

\bibitem{marianna} Safronova MS,  Porsev SG,  Safronova UI,  Kozlov MG and  Clark CW 2013 Blackbody-radiation shift in the Sr optical atomic clock {\em Phys. Rev. A} \textbf{87}, 012509

\bibitem{porsev2}  Porsev SG and  Derevianko A 2006 Multipolar theory of blackbody radiation shift of atomic energy levels and its implications for optical lattice clocks {\em Phys. Rev. A} \textbf{74}, 020502(R) 

\bibitem{chem}  Sansonetti JE and  Nave G 2010 Wavelengths, Transition Probabilities, and Energy Levels for the Spectrum of Neutral Strontium (Sr I) {\em J. Phys. Chem. Ref. Data} \textbf{39}, 033103

\bibitem{Lisdat09}  Lisdat C, Vellore Winfred JSR,  Middelmann T,  Riehle F and  Sterr U 2009 Collisional Losses, Decoherence, and Frequency Shifts in Optical Lattice Clocks with Bosons {\em Phys. Rev. Lett.} {\bf 103} 090801 

\bibitem{Morzynski13} Morzynski P, Wcislo P, Ablewski P, Gartman R, Gawlik W, Maslowski P, Nagorny B, Ozimek F, Radzewicz C, Witkowski M, Ciurylo R and  Zawada M 2013  Absolute frequency measurement of rubidium 5S-7S two-photon transitions, {\em Opt. Lett.} {\bf 38}  4581

\end{thebibliography}

\end{document}